\documentclass[titlepage,12pt]{utarticle}
\usepackage{bbm}
\usepackage{amsmath}
\usepackage{amsfonts}
\numberwithin{equation}{section}

\usepackage{graphicx}
\usepackage{amsmath}
\usepackage{color}
\usepackage{multirow}

\def\br{\mathbf r}
\def\bs{\mathbf s}
\def\bg{\mathbf g}
\def\bx{\mathbf x}

\begin{document}
\title{Exact single-sided inverse scattering in three-dimensions}
\author{Harun Omer}
\oneaddress{\tt firstname.lastname@kaust.edu.sa \tt firstname@mit.edu}
    %{\tt email}\\{~}\\
 \nobreak
\Abstract{During the past three years, Wapenaar, Snieder, Broggini and others have developed an algorithm to compute
the Green's function for any point inside a medium to points on the surface from measurements on that surface only. Their algorithm
is based on focusing an incoming wavefield to a single point in order to create a virtual source at the focus.
The procedure has been justified only by heuristic arguments. In this paper I am using simple physical arguments to prove an integral equation for single-sided, higher-dimensional inverse scattering. This integral equation is equivalent to the Wapenaar iteration algorithm. The equation
will be exact, including all internal multiple reflections. The derivation makes use of time-invariance but does not use the explicit form of the wave equation.
It is therefore not only applicable to the acoustic wave equation, but also to other time-reversal invariant systems. Potential areas
include seismology, electronics, microwave and ultrasonic inverse scattering as well as quantum physics. The simplicity and generality of the argument will make
this paper accessible to researchers from all the mentioned fields.}
\maketitle
%%%%%%%%%%%%%%%%%%%%%%%%%%%%%%%%%%%%%%%%%%%%%%%%%%%%%%%%%%%%%%%%%%%%%%%%%%%%%%%%%%%%%%%%%%%%%%%%%%%%%%%%%%
\section{Introduction}
Recently Waapenar, Snieder, Broggini and others dug out work by Rose~\cite{Rose1,Rose2,Rose3} on one-dimensional
inverse scattering, extended it somewhat to be able to read off the Green's function and generalized it to three dimensions~\cite{Wapenaar1,Wapenaar2,Wapenaar3,Wapenaar5}.
The idea is as follows. Given sources and receivers on the surface of a medium, a shot gather is recorded.
Using the data of the shot gather, one iteratively constructs an
incoming wavefield emitted at the sources, which focuses at a time $t=0$ entirely at a single point $\mathbf{x}$ in the subsurface.
'Focusing' means that at a specific point in time, the wavefield vanishes everywhere except at $\mathbf{x}$.
Since the basic acoustic wave equation is invariant under time-reversal, knowing the focusing wavefield for some point $\bx$ amounts to
knowing a wave originating at that point and propagating backwards towards the sources. This is the Green's function
of that point, called the virtual source. The appeal is that apart from noise and numerical inaccuracies, the method
is exact and takes into account all internal reflections. A velocity model is only needed to determine the position of
the virtual source but not as an input to the calculation of the Green's function. The Green's function for every point in an acoustic
medium encodes its entire information and can for example also used for seismic imaging as was done in~\cite{Behura}.
The iterative algorithm for 1D had been developed in~\cite{Rose1} and amounts to solving the Gelfand-Levitan~\cite{Rose2,Burridge} equation --
also called Marchenko equation or Gelfand-Levitan-Marchenko equation. In~\cite{Wapenaar1} the theory was applied to 1D synthetic data
and in~\cite{Wapenaar3} the aforementioned iteration algorithm has been generalized to three dimensions and justified by heuristic arguments.
Having a more rigorously proven three-dimensional analog to the Gelfand-Levitan equation would be beneficial for a number of potential applications
in areas such as seismic interferomety, seismic imaging, inverse scattering of microwaves, of ultrasonic waves and possibly scattering in quantum mechanics. The purpose of this paper is to obtain the exact equation for single-sided inverse scattering from simple physical arguments and very basic math. The
resulting equation is equivalent to the aforementioned iteration scheme. Furthermore, since inverse scattering equations have been used in a variety of areas and contexts, a literature search is at times confusing. I intend to clear some of the fog by gathering the known scattering equations and their application in the last section of this paper.
%%%%%%%%%%%%%%%%%%%%%%%%%%%%%%%%%%%%%%%%%%%%%%%%%%%%%%%%%%%%%%%%%%%%%%%%%%%%%%%%%%%%%%%%%%%%%%%%%%%%%%%%%%
%%%%%%%%%%%%%%%%%%%%%%%%%%%%%%%%%%%%%%%%%%%%%%%%%%%%%%%%%%%%%%%%%%%%%%%%%%%%%%%%%%%%%%%%%%%%%%%%%%%%%%%%%%
\section{Derivation of the inverse scattering equation}
The reflection of a wave incident on a medium is a wavefield $R(x,t)$ propagating in the opposite direction to the incident wave. In practice, the wave is recorded at a receiver at a fixed position and we assume that beyond the receiver it keeps propagating freely so we can write $R(x+t)$ just from measurements at the receiver. 
In one dimension, we assume the source and receiver to coincide. In higher dimensions, a source $\bs \in S$ anywhere on the medium surface $S$ causes a scattered signal to arrive at all receivers $\bg \in S$. Therefore the reflection response in a higher-dimensional setting has one more parameter, $R(\bg,\bs,t)$. In geophysics this is the shot gather which is recorded in seismic experiments.\\
Physically speaking, the Gelfand-Levitan equation connects the reflection data from the scattering problem to the wavefield of a focusing wave.
Numerous derivations of the 1D Gelfand-Levitan equation for inverse scattering already exist, for example~\cite{Rose2,Rose3,Burridge}. Here the equation
is rederived in a way which allows generalization to higher dimension. The basic idea is simple: We write down the wavefields of both the reflection problem
as well as the focusing wave. Since the wave equation is linear, we can superimpose solutions, including
solutions shifted in time. We superimpose focusing waves in such a way that the resulting wavefield is brought in agreement with the scattering problem. The
result will be the desired integral expression which we can solve for the unknowns. That's all there is to it.
\begin{table}
\begin{tabular}{l|l}
\multicolumn{2}{l}{\textbf{(a) Wavefield for the reflection problem:}} \\
$ x \not\in M$ & $x \in M$\\
\hline
$\delta(x-t) + R(x+t)$ & $K(x,t)$ \;\;(with $K(x,t) = 0 \mbox{ for } t < x.$) \\
\multicolumn{2}{l}{}\\
\multicolumn{2}{l}{\textbf{(b) Wavefield of a focusing wave:}} \\
$ x \not\in M$ & $x \in M$\\
\hline
$\delta(x-t)$ & $\delta(x-t) + g^s(x,t)$\\
\multicolumn{2}{l}{}\\
\multicolumn{2}{l}{\textbf{(c) Wavefield of the reflection problem in terms of focusing waves:}} \\
$ x \not\in M$ & $x \in M$\\
\hline
\multirow{2}{*}{$\underbrace{\displaystyle \int_{-\infty}^{x}\delta(x-t') \left[\delta(t-t') + R(t+t')\right] dt'}_{=\delta(x-t) + R(x+t)}$} & 
$\displaystyle \displaystyle \int_{-\infty}^{x}\delta(x-t') \left[\delta(t-t') + R(t+t')\right] dt'$\\
&$\displaystyle + \int_{-\infty}^{x}g^s(x,t') \left[\delta(t-t') + R(t+t')\right]dt'$
\end{tabular}
\caption{In (c) we superimpose focusing waves in such a way that the field is identical to (a) in the region $ x \not\in M$.}
\label{tab:GelfandLevitan}
\end{table}
\subsection{Derivation in 1D}
In our setup of the 1D scattering problem, an impulse signal $\delta(x-t)$ is incident from the left on a
medium $M$. One part of the wave is reflected, $R(x+t)$, while another, as of yet unknown part $K(x,t)$, enters the medium. 
Typically the reflected wave can be measured while the field $K(x,t)$ is unknown. Causality and the finite speed of wave propagation require that the amplitude must vanish before the direct arrivals of the transmitted component of the wave:
\begin{eqnarray}
K(x,t) = 0 \mbox{ for } t < x. \label{eq:vaneq}
\end{eqnarray}
What we know about the wavefield is summarized in Table~\ref{tab:GelfandLevitan}a.
We now want to relate this equation to a focusing wave. A focusing wave is an incoming impulse wave incident on the medium from the left in our setup. Inside the medium
it is followed by a 'tail' in such a way that at the focusing time $t=0$ the wave amplitude vanishes everywhere except at the focal point,
where it gives rise to a $\delta$-function peak. Such a cancellation of amplitudes rules out that reflected waves propagate to the left since otherwise they would keep propagating freely outside the medium even at focusing time. The wavefield of a focusing wave therefore has the form shown in Table~\ref{tab:GelfandLevitan}b.
The unknown field in the medium is denoted by $g^s(x,t)$. We next want to superimpose such waves so that they give rise to the wavefield of the reflection problem outside the medium. This is done in Table~\ref{tab:GelfandLevitan}c. For $x \not\in M$ ~\ref{tab:GelfandLevitan}a and~\ref{tab:GelfandLevitan}c are identical.\footnote{Strictly speaking, the product of delta distributions in the derivation is a delicate issue, but in real-world situations the incoming wave cannot be a delta-distribution anyway. To derive this more rigorously, one would convolve the equations with an arbitrary source wavelet before superimposing. Then $\delta(x-t')$ would be replaced with the convolved wavelet while $\delta(t-t')$ remains a delta function. Since the wavelet is arbitrary, it can afterwards be dropped again.}
%%%%%%%%%%%%%%%%%%%%%%%%%%%%%%%%%%%%%%%%%%
Since the wave propagation solution is unique, the waves inside the medium must also be identical.
Using eq.~(\ref{eq:vaneq}), we have:
\begin{eqnarray}
\delta(x-t) + R(t+x) + \int_{-\infty}^{x}g^s(x,t') \left[\delta(t-t') + R(t+t')\right]dt' = 0 \mbox{ for } t < x.
\end{eqnarray}
which within the domain of validity is equivalent to,
\begin{eqnarray}
g^s(x,t) + R(t+x) + \int_{-\infty}^{x}g^s(x,t') R(t+t') dt' = 0 \mbox{ for } t < x.\label{eq:GLeq}
\end{eqnarray}
The reason why the upper bound of the integral is at a finite $x$ deserves explanation: The bound has to be {\it at least} $x$ so the waves are in agreement in $x\not\in M$. It also has to be {\it at most} $x$ to satisfy the constraint at $t < x$ in $x \in M$. 
The resulting equation~(\ref{eq:GLeq}) is the Gelfand-Levitan integral equation which allows solving for the focusing wavefield for a given reflection response $R(t)$.
%%%%%%%%%%%%%%%%%%%%%%%%%%%%%%%%%%%%%%%%%%%%%%%%%%%%%%%%%%%%%%%%%%%%%%%%%%%%%%%%%%%%%%%%%%%%%%%%%%%%%%%%%%
\subsection{Derivation in higher dimensions}
\begin{figure}
\includegraphics[width=50mm]{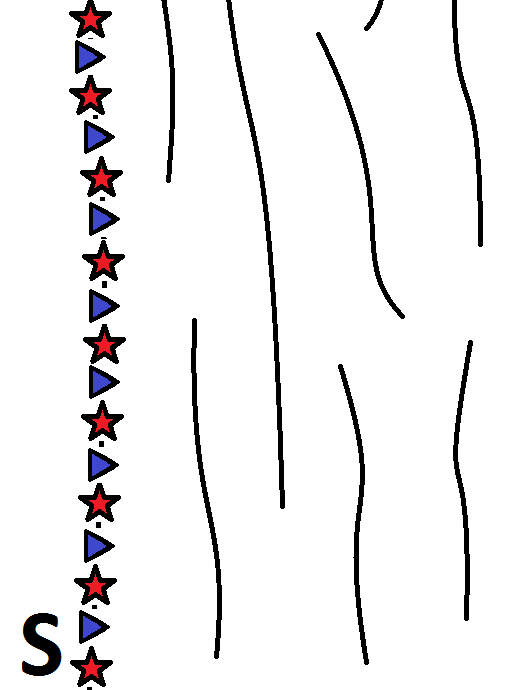}
\includegraphics[width=50mm]{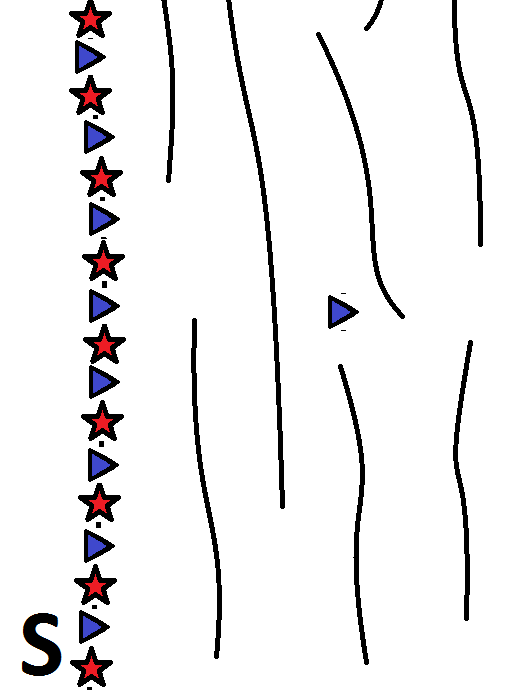}
\includegraphics[width=50mm]{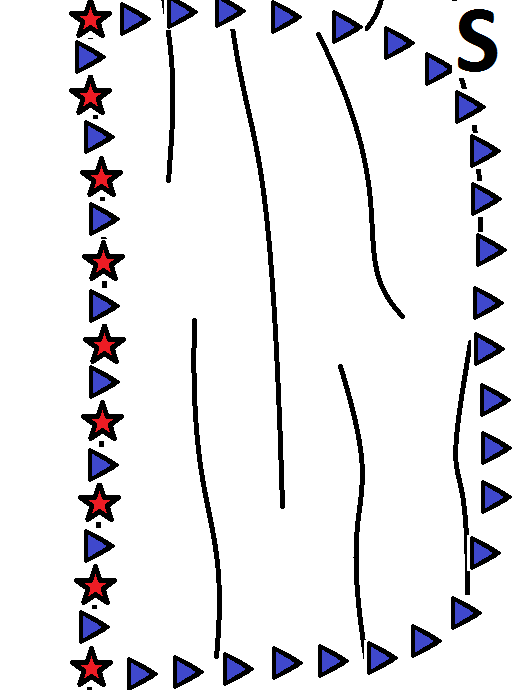}
\caption{A schematic experimental setup with sources and receivers on the surface $S$. With the proposed method (l.) one-sided surface measurements are sufficient.
Seismic interferometry (m.) needs a source inside the medium for each subsurface point of interest. The Newton-Marchenko equation requires receivers surrounding the entire medium.}
\label{fig:setup}
\end{figure}

The derivation for higher dimensions proceeds along the same lines as in 1D. The incident wave now enters through a surface $S$ bounding the
medium from one side as shown in Fig.~\ref{fig:setup}.
In principle the reflection problem can be generalized in different ways. The incoming wave can be an impulse wave incident at only
one point on the surface $S$. This is a typical situation arising in quantum scattering and equations for a generalized Gelfand-Levitan equation
have been derived for example in~\cite{Newton4}. Particularly in seismic experiments, one deals with more than one point of incidence.
Reflection responses are gathered all over a surface region $S$ from shots all over that same region $S$. The reflection response $R(\bg,\bs,t)$ -- the shot gather -- therefore has two parameters: The position of the source $\bs \in S$ and the receiver $\bg \in S$ in contrast to the one-dimensional case.
This allows us to illuminate the medium from the entire surface $S$ instead of from only one point.
The incoming direct wave, which was a delta-function in 1D, can therefore be generalized to an incoming wavefront
passing through the entire surface $S$. In analogy to the 1D wave which focuses at $(x,t)=(0,0)$ if it were to continue propagating freely, 
the Green's function $G^d(x,-t)$ also focuses at the origin. The negative $t$ is because it is the time reverse of the usual Green's function $G^d(x,t)$, describing the propagation from a point source at the origin to $\bx$, which is the solution of,
\begin{eqnarray}
L \cdot G^d(\bx,t) = \delta(\bx) \delta(t),
\end{eqnarray}
where $L$ is the operator of the wave equation. Note that we are not using its explicit form anywhere so results will be valid for all systems
invariant under time reversal. When the wave is a time-reversed Green's function, the scattering field takes the form given in Table~\ref{tab:3d}(a).
The focusing wave is given in~\ref{tab:3d}(b). And in~\ref{tab:3d}(c) we again take a superposition of focusing waves so that they sum up to the
wavefield of the scattering problem. In the 1D case, the upper bound of the integral was $x$, assuming that the wave propagates with speed $c=1$.
We now have waves originating from all over $S$, and the integral bound is given by the first arrivals $t_f(\bx,\bs)$ from the source $\bs$ to $\bx$.
This is also a generalization from before, since we no longer have to assume constant speed of propagation. Apart from this difference, the argument
proceeds as before and we compare~\ref{tab:3d}(a) and~\ref{tab:3d}(c) in $x \in M$. For $t<t_f$ one obtains,
\begin{eqnarray}
\begin{array}{l}
\boxed{
\displaystyle  G^s(\bx,-t) \!+\!\int_S \int_{-\infty}^{t_f} \!\!\! R(\bx,\bs,t-t')G^d(\bs,-t')dt' d\bs \!+\! \int_S \! \int_{-\infty}^{t_f} \!\!\! R(\bx,\bs,t-t')G^s(\bs,-t)dt'd\bs=0.}\label{eq:generalization}
\end{array}
\end{eqnarray}
Since this equation is valid only for $t < t_f$, the direct wave $G^d(\bx,-t)$ does not contribute anything and has been dropped from the equation.
This is in analogy to the dropped $\delta(t-x)$ in 1D. Equation~(\ref{eq:generalization}) is the 3D generalization of the Gelfand-Levitan equation.
%%%%%%%%%%%%%%%%%%%%%%%%%%%%%%%%%%%%%%%%%%%%%%%%%%%%%%%%%%%%%%%%%%%%%%%%%%%%%%%%%%%%%%%%%%%%%%%%%%%%%%%%%%
\begin{table}
\begin{tabular}{l|l}
\multicolumn{2}{l}{\textbf{(a) Wavefield for the reflection problem:}} \\
$ \bx \not\in M$ & $\bx \in M$\\
\hline
$\displaystyle G^d(\bx,-t) \!\!+ \! \int_S \! \int_{-\infty}^{t_f}\!\!\!\!\! R(\bx,\bs,t-t') G^d(\bs,-t')dt' d\bs$ & $K(\bx,t)$ with $K(\bx,t)=0$ for $t < t_f$\\
\multicolumn{2}{l}{}\\
\multicolumn{2}{l}{\textbf{(b) Wavefield of a focusing wave:}} \\
$ \bx \not\in M$ & $\bx \in M$\\
\hline
$\displaystyle G^d(\bx,-t)$ & $G^d(\bx,-t) + G^s(\bx,-t)$\\
\multicolumn{2}{l}{}\\
\multicolumn{2}{l}{\textbf{(c) Wavefield of the reflection problem in terms of focusing waves:}} \\
$\bx \not\in M$ & $\bx \in M$\\
\hline
\multirow{2}{*}{\resizebox{.5\hsize}{!}{$\underbrace{\displaystyle \int_S \! \int_{-\infty}^{t_f} \!\!\! G^d(\bs,-t')\left[\delta(\bx -\bs) \delta(t-t') + R(\bx,\bs,t-t')\right]dt'd\bs}_{=\displaystyle G^d(\bx,-t) + \int_S \int_{-\infty}^{t_f} R(\bx,\bs,t-t') \cdot G^d(\bs,-t')dt' d\bs}$}} &
\resizebox{.5\hsize}{!}{
$\displaystyle \! \int_S \! \int_{-\infty}^{t_f} \!\!\! G^d(\bs,-t')\left[\delta(\bx -\bs) \delta(t-t') + R(\bx,\bs,t-t')\right]dt'd\bs$}\\&\!\!+\!\!
\resizebox{.5\hsize}{!}{
$\displaystyle \! \int_S \! \int_{-\infty}^{t_f} \!\!\! G^s(\bs,-t')\left[\delta(\bx -\bs) \delta(t-t') \!+\! R(\bx,\bs,t-t')\right]dt'd\bs$}
\\
\multicolumn{2}{l}{}\\
\end{tabular}
\caption{In (c) we again superimpose focusing waves in such a way that the field is identical to (a) in the region $ \bx \not\in M$.}
\label{tab:3d}
\end{table}
%%%%%%%%%%%%%%%%%%%%%%%%%%%%%%%%%%%%%%%%%%%%%%%%%%%%%%%%%%%%%%%%%%%%%%%%%%%%%%%%%%%%%%%%%%%%%%%%%%%%%%%%%%
%%%%%%%%%%%%%%%%%%%%%%%%%%%%%%%%%%%%%%%%%%%%%%%%%%%%%%%%%%%%%%%%%%%%%%%%%%%%%%%%%%%%%%%%%%%%%%%%%%%%%%%%%%
\section{Proof of the iteration algorithm}
\subsection{Integral equation from iteration}
We now derive the integral equation~(\ref{eq:generalization}) from the iterative algorithm of Wapenaar et al.~\cite{Wapenaar3}.
The algorithm relates downgoing waves $p^+ ( \bg ,t )$ and upgoing waves $p^- (\bg , t )$ through the following iteration,
\begin{equation}
\label{Iter.1new}
p_k ^+ ( \bg , -t ) = G^d( \bg , t ) - w (\bg , t) p^-_{k-1} (\bg , t )\;,
\end{equation}
\begin{equation}
\label{Iter.2new}
p_k ^- ( \bg , t ) = \int_S \int_{-\infty}^{\infty} R ( \bg , \bs , t-t' ) p_k ^+ (\bs , t' ) dt' d \bs \;.
\end{equation}
The window function $w (\bg , t)$ satisfies,
\begin{equation}
\label{Iter.3new}
w (\bg , t) = \left\{
\begin{array}{l}
1\makebox{\hspace{0.8cm}for\hspace{0.8cm}}|t|< t_f (\bg , \br )\\
0\makebox{\hspace{0.8cm}for\hspace{0.8cm}}|t|> t_f (\bg , \br )
\end{array}
\right.
\end{equation}
At the point of convergence, we can insert eq.~(\ref{Iter.1new}) into eq.~(\ref{Iter.2new}) and obtain an integral equation for $p^-(\bg,t)$,
\begin{equation}
\begin{array}{rcl}
p^- ( \bg , t ) \!&=&\!
\displaystyle \!\int_S\! \int_{\infty}^{\infty}\!\!\! R ( \bg , \bs , t -t') G^d(\bs , -t' ) dt' d\bs
-\int_S \int_{\infty}^{\infty} R ( \bg , \bs ,t-t') w(\bs,-t') p^- (\bs , -t') dt' d\bs\\
&=&
\displaystyle \int_S \int_{-\infty}^{\infty} R ( \bg , \bs ,t-t') G^d(\bs ,-t') dt' d \bs
-\int_S \int_{-t_f (\bs,\br)}^{t_f (\bs,\br)} R ( \bg , \bs ,t-t') p^- (\bs ,-t') dt' d \bs
\end{array}\label{integraleq}
\end{equation}
We can identify,
\begin{eqnarray}
G^s(\bs,t) \equiv - p_0^-(\bs,t),
\end{eqnarray}
and recover eq.~(\ref{eq:generalization}),
\begin{equation}
%\begin{array}{rcl}
0= \! G^s( \bg , t )+\!
\int_S \int_{-\infty}^{\infty}\!\!\! R ( \bg , \bs ,t-t') G^d(\bs ,-t') dt' d \bs
+\!\int_S \int_{-t_f (\bs,\br)}^{t_f (\bs,\br)}\!\!\! R ( \bg , \bs ,t-t') G^s(\bs ,-t') dt' d \bs.
%\end{array}
\label{eq:integraleq2}
\end{equation}
The integral bounds seemingly differ from eq.~(\ref{eq:generalization}), but in the respective regions the contribution from the integral vanishes.
The connection of the iteration algorithm with eq.~(\ref{eq:integraleq2}) has already been proposed in~\cite{Wapenaar3} in eq. (8)-(9) and by~\cite{Roel}.
%%%%%%%%%%%%%%%%%%%%%%%%%%%%%%%%%%%%%%%%%%%%%%%%%%%%%%%%%%%%%%%%%%%%%%%%%%%%%%%%%%%%%%%%%%%%%%%%%%%%%%%%%%
\subsection{Iteration from integral equation}
One way~\cite{merab} to solve an equation of the form of eq.~(\ref{eq:generalization}) is by introducing an auxiliary parameter $\lambda$ which we later set to unity:
\begin{equation}
\begin{array}{rcl}
G^s ( \bg , t ) &=&
\displaystyle -\int_S \int R ( \bg , \bs , t - t') G^d (\bs ,-t') dt'd \bs
-\lambda \int_S \int_{-t_f (\bs,\br)}^{t_f (\bs,\br)} R ( \bg , \bs , t - t') G^s (\bs , -t') dt' d \bs\\
\end{array}\nonumber
\end{equation}
Writing $G^s(\bg, t)$ as a power series,
\begin{eqnarray}
G^s ( \bg ,t) = \displaystyle \sum_{i=0}^{\infty} \lambda^i G^s_{i+1}(\bg,t),
\end{eqnarray}
we can solve for each order of $\lambda$ and obtain for $k \in \mathbb{N}_0$,
\begin{eqnarray}
\begin{array}{rcl}
%G^s_0(\bg,t) \equiv G^d(\bg,t)\\
%G^s_1 (\bg, t) &=& \displaystyle -\int_S \int R ( \bg , \bs , t - t') G^d (\bs , -t') dt'd \bs\\
G^s_{k+1} (\bg,t) &=& \displaystyle -\int_S \int_{-t_f (\bs,\br)}^{t_f (\bs,\br)} R ( \bg , \bs , t - t') G^s_{k} (\bs , -t') dt' d \bs,
\end{array}\label{eq:iteration}
\end{eqnarray}
where the starting value of the iteration is set to,
\begin{eqnarray}
%\begin{array}{rcl}
G^s_0(\bg,t) \equiv G^d(\bg,t).
%\end{array}
\end{eqnarray}
This is equivalent to the Wapenaar algorithm. The Banach fixed point theorem ensures that the integral equation has
a unique solution and that the series converges for physically reasonable shot gathers.\\
%%%%%%%%%%%%%%%%%%%%%%%%%%%%%%%%%%%%%%%%%%%%%%%%%%%%%%%%%%%%%%%%%%%%%%%%%%%%%%%%%%%%%%%%%%%%%%%%%%%%%%%%%%
Before I conclude this section, let me briefly remark on noise. The structure of the recursion eq.~(\ref{eq:iteration}) as well as the iteration algorithm work is such that they provide an intuitive understanding of the noise to be expected in the data. 
Suppose that the reflection response $R ( \bg , \bs , t )$ has been measured from an 
experiment where the incoming wave is a bandlimited source-wavelet $s(t)$ -- typically a Ricker function instead of an ideal $\delta$-distribution. If we do not deconvolve $R ( \bg , \bs )$, an error will be picked up at every iteration step. At every iteration, the new contribution would be convolved with the source
wavelet $s(t)$. The $k$-th contribution $G^s_{k} (\bg , t)$ is convolved $k$-times with the source wavelet. In reality one should of course deconvolve $R( \bg , \bs , t )$, but for a noisy signal deconvolution is never exact. A wave-front normally narrows down a bit after deconvolution but a sharp $\delta$ peak will never be reached. Therefore qualitatively the argument remains true. In all cited papers on the subject, synthetic data has been used where deconvolution problems do not occur. Another way to think of noise is by realizing that the contribution at the $k$-th level comes roughly speaking comes from a wave which has propagated in the medium $k$ times as long as the wave at iteration $1$.
%%%%%%%%%%%%%%%%%%%%%%%%%%%%%%%%%%%%%%%%%%%%%%%%%%%%
\section{A tour on scattering equations and their application}
A number of similar equations exist in the literature and have been applied to various scattering problems.
It is worthwhile clearing up some confusion here by listing the main equations and their purpose. The one-dimensional
Newton-Marchenko equation reads,
\begin{eqnarray}
\begin{array}{rcl}
g^s(e,x,t) &=& \displaystyle \sum_{e'=-1,1} R(e,e',t+e' x) + \sum_{e'=-1,1}\int_{-\infty}^{\infty} R(e,e',t'+e' x) g^s(e',x,t') dt'
\end{array}\label{eq:NM1d}
\end{eqnarray}
where $g^s(e,x,t)=0$ for $t<ex$. The parameter $e = \pm 1$ distinguishes between the direction of wave propagation. Rose initially used this equation in his work 
and showed that it solves for the field $g^s(e,x,t)$ that focuses to $\delta(x)$ at $t=0$~\cite{Rose1}.
Its 3D generalization is given in slightly different form in~\cite{Newton1} and can be written,
\begin{eqnarray}
g^s(\mathbf{e}_i,t_f,t) = \int_{S^2} R(\mathbf{e}_i,\mathbf{e}_s,t+t_f)d \mathbf{e}_s + \int_{S^2} \int_{\infty}^{\infty} R(\mathbf{e}_i,\mathbf{e}',t+t') g^s(-\mathbf{e}',t_f,t') d\mathbf{e}' dt'.
\label{eq:NM3d}
\end{eqnarray}
where $g^s(\mathbf{e}_i,t_f,t)$ is the scattered field for an impulse injected in direction $\textbf{e}_i$. In eq.~(\ref{eq:NM1d}) the reflection response is recorded on both sides of the medium, $e'=1$ and $e'=-1$. In other words, the equation depends on the reflected as well as the transmitted wave. In eq.~(\ref{eq:NM3d}) the summing over both sides has been replaced by an integral over a closed surface $S^2$ surrounding the entire medium. The scattered wavefield is recorded all over the closed surface. These equations are known from scattering theory, where they have been applied to the plasma equation.
In momentum space it reads,
\begin{eqnarray}
(\Delta + k^2 - V(x))u(k,x)=0.
\label{eq:plasma}
\end{eqnarray}
Note also that in the scattering problem an impulse signal (or in momentum space a plane wave) is scattered as it passes through a medium and the scattering response is recorded. This wave is incident from one specific direction only, here $\mathbf{e}_i$. From the theoretical point of view it is of course attractive that a simple plane wave is all that is needed to solve the scattering problem, but in the presence of noise it is preferable to have an incident wavefield coming from an entire area covered with sources. Furthermore, for seismic experiments it is impractical to bury sources or receivers in the ground. We want to deduce the properties of the subsurface from measurements on the surface only. In other words, we are looking for an equation using one-sided illumination only.
Rose has solved that problem in 1D using focusing waves. In 1D the incident wave would be a delta function impulse
followed by a tail such that all scattered waves are cancelled at the focusing time. It was shown that an equation called Marchenko equation or Gelfand-Levitan equation solves the problem~\cite{Rose2,Rose3}. While for example in~\cite{Behura} the authors refer to Newton and Marchenko by dubbing their imaging method Newton-Marchenko-Rosen imaging, their single-sided illumination algorithm in 1D actually corresponds to the Gel'fand-Levitan equation. It has been rederived in
this paper and is given in eq.~(\ref{eq:GLeq}). The Newton-Marchenko and the Gel'fand-Levitan equations might look similar but they are rather different. The former needs illumination from all sides, the time integral is unbounded and the solution is the entire wave-field. The latter needs illumination only from one side, the integration ends at the time of the first arrivals and its solution is only the scattered field. 
The equations appear in the literature in somewhat varying forms and also for different applications.
Characteristic of Marchenko-type equations are integral limits of only $\pm \infty$ or $0$ in the time integration whereas Gelfand-Levitan equations have one integration bound at a different value. A "Generalized Gelfand-Levitan" equation also exists in the literature~\cite{Newton1}.
Like all Gelfand-Levitan type equations it has a finite bound of integration over the time variable, but it also requires an integration over a closed surface around
the medium, so it is not a single-sided scattering equation. Its integration kernel is also different from the ones in the table.
To avoid causing confusion, I therefore abstained from calling the equation in Table~\ref{tab:scateq} generalized Gelfand-Levitan equation or 3D Gelfand-Levitan equation. 
\begin{table}
\begin{tabular}{p{8cm}cccc}
Common name & Eqn & Dim & recording surface\\
\hline\hline
Newton-Marchenko & (\ref{eq:NM1d}) & 1D & both sides\\
3D Newton-Marchenko & (\ref{eq:NM3d}) & 3D & closed surface\\
Gelfand-Levitan {\it or} Marchenko & (\ref{eq:GLeq}) & 1D & single-sided\\
N/A & (\ref{eq:generalization}) & 3D & single-sided
\end{tabular}\label{tab:scateq}
\end{table}
%%%%%%%%%%%%%%%%%%%%%%%%%%%%%%%%%%%%%%%%%%%%%%%%%%%%%%%%%%%%%%%%%%%%%%%%%%%%%%%%%%%%%%%%%%%%%%%%%%%%%%%%%%
\section{Summary and conclusion}
I have derived an integral equation equivalent to the Wapenaar iteration algorithm from simple physical reasoning only. The equivalence with the iteration algorithm
was proven. The formula can be applied for different purposes. One can find a Green's function for a 'virtual source' under the subsurface, which is nothing but
the time-reversal of the wavefield of the focusing wave for $t > t_f$~\cite{Wapenaar1,Wapenaar3}. The advantage is that it has been recovered without the need to
place a source or receiver inside the medium. From the fundamental identity (dubbed by Newton as "miracle equation")~\cite{Newton1} the potential $V(\bx)$ of the plasma wave equation~(\ref{eq:plasma}) can be recovered. And by separating the up-going from the down-going wave, the Green's function can be used as the input of an imaging condition and provide an image with all internal multiples. Examples with synthetic data are shown in~\cite{Behura}. Using the integral equation, instead of the Wapenaar algorithm there are presumably more effective ways to compute images, perhaps using a layer-peeling algorithm. A variety of applications are thinkable since knowing the Green's function for every point inside a medium amounts to knowing everything about the scattering problem.
%Since for example the equations belong to the Wiener-Hopf equations, it is also known how to image noisy data and obtain a least-squares solution. 
%%%%%%%%%%%%%%%%%%%%%%%%%%%%%%%%%%%%%%%%%%%%%%%%%%%%%%%%%%%%%%%%%%%%%%%%%%%%%%%%%%%%%%%%%%%%%%%%%%%%%%%%%%
\section{Acknowledgements}
I am grateful to Jerry Schuster, Filippo Broggini and Roel Snieder for useful discussions. Special thanks to Alison Malcolm for many helpful comments
on the manuscript. I also acknowledge the generous support of KAUST and the CSIM sponsors (http://csim.kaust.edu.sa) and thank the KAUST Supercomputing Lab for the computer cycles they donated to our research. We are especially grateful for the use of the SHAHEEN supercomputer.
%%%%%%%%%%%%%%%%%%%%%%%%%%%%%%%%%%%%%%%%%%%%%%%%%%%%%%%%%%%%%%%%%%%%%%%%%%%%%%%%%%%%%%%%%%%%%%%%%%%%%%%%%%

\end{document}